\documentclass[twocolumn,letterpaper,amsmath,amssymb,floatfix,aps,superscriptaddress]{revtex4} 

\usepackage{graphicx}
\usepackage{dcolumn}
\usepackage{bm}
\usepackage{epsfig}

\newcommand{\Av}[1]{{\bf #1}}

\def\ln{{\operatorname{ln}}}

\begin{document}
\preprint{ }

\title{Counterion-mediated Electrostatic Interactions between Helical Molecules }
\author{M. Kandu\v c$^{1}$, J. Dobnikar$^{1}$  and R. Podgornik$^{1,2}$\\
$^1$ Department of Theoretical Physics, \\
J. Stefan Institute, SI-1000 Ljubljana, Slovenia\\
$^2$ Department of Physics, Faculty of Mathematics and Physics, \\
University of Ljubljana, SI-1000 Ljubljana, Slovenia}

\begin{abstract} We study the interaction of two cylinders with helical charge distribution mediated by neutralizing counterions, by analyzing the separation as well as the azimuthal angle dependence of the interaction force in the weak and strong coupling limit. While the azimuthal dependence of the interaction in the weak coupling limit is overall small and mostly negligible, the strong coupling limit leads to qualitatively new features of the interaction, among others also to an orientationally dependent optimal configuration that is driven by angular dependence of the correlation attraction. We investigate the properties of this azimuthal ordering in detail and compare it to existing results.

\end{abstract}

\maketitle

\section{Introduction}
Electrostatic interactions often play a dominant role in biological and soft-matter systems \cite{holm}. In aqueous environments charges on macromolecular surfaces such as membranes, self-assembled micelles, globular proteins and fibrous polysaccharides tend to dissociate and affect a wealth of functional, structural and dynamical properties \cite{Andelman}.
Charged biological macromolecules and self-assembled colloidal systems also show a ubiquitous patterning of the structural charges that are seldom distributed uniformly on the macromolecular surfaces. This can be most clearly seen in the patchy nature of charge distribution on various proteins, the notorious double-helical charge motif seen on DNA \cite{Kornyshevrev} and then all the way to self-assembled mixed charged bilayers \cite{perkin} and decorated carbon nanotubes (CNT). In the latter case too, as was realized recently  \cite{JagotaNat, JagotaSci, JagotaLustig}, one can have a pronounced helical motif of the charges when CNTs associate with single stranded DNAs that charges it up and makes this quintessential hydrophobic macromolecule soluble in water.

Interactions between macromolecules exhibiting helical charge motifs were investigated vigorously by Kornyshev and Leikin \cite{Kornyshev} who formulated a Debye-H\" uckel-Bjerrum type theory for interactions between DNAs. In their approach the charge distribution on DNAs, including the counterions in close proximity to the charged phosphate groups, is assumed {\sl a priori} and the interaction between two such molecules in close proximity is calculated within the Debye - H\" uckel approximation based on this assumption. In the case of ss-DNA covered CNTs there were various attempts to derive the presumed helical charge motif based on either approximate Debye - H\" uckel (DH) arguments \cite{JagotaLustig} or MD simulations \cite{JagotaManohar}. Here the interactions between decorated CNTs are less well studied since it is not clear whether the helical charge motif would not be significantly perturbed by the interaction itself. However, in both cases of macromolecules with helical charge motifs, the electrostatic interaction is basically treated on the DH level and should thus be valid asymptotically for sufficiently large separations between macromolecules \cite{olli}. There are other regions in the parameter space of these type of systems \cite{hoda} where one expects the DH framework to break down and one should see the emergence of completely different type of physics. Unfortunately we are still not close to having a consistent and possibly asymptotically exact formulation of the statistical mechanics of charged systems, containing a mixture of salts, polyvalent counterions and strong fixed charges, though there is some recent progress in this direction \cite{olli}.

In the absence of a general approach that would cover thoroughly all the regions of the parameter space one has to take recourse
to various partial formulations that take into account only this or that facet of the problem. In this respect the counterion-only or the one-component Coulomb fluid model system has proved to be of substantial value \cite{hoda}. Heuristically as well as numerically.  A proper understanding of the behavior of charged systems would thus start with  the analysis of counter-ion distribution around charged macromolecular surfaces, neglecting completely the effects of salt. The traditional approach to these one-component Coulomb fluids has been the mean-field Poisson-Boltzmann (PB) formalism applicable at weak surface charges, low counter-ion valency and high temperature \cite{Oosawa, Ohnishi, Naji}. The limitations of this approach become practically important in highly-charged systems where  counterion-mediated interactions between charged bodies start to deviate substantially from the mean-field accepted wisdom \cite{hoda}. One of the most important recent advances in this field has been the systematization of these non-PB effects based on the notions of {\sl weak} and {\sl strong} coupling approximations. This approach has been pioneered by Rouzina and Bloomfield \cite{Rouzina}  and elaborated later by Shklovskii {\em et al.} \cite{shklovskii}, Levin {\em et al.} \cite{Levin},  and brought into final form by Netz {\em et al.} \cite{Netz,hoda,Naji}. These two approximations allow for an explicit and exact treatment of charged systems at two disjoint limiting conditions whereas the parameter space in between can be analysed only approximately and is mostly accessible solely {\sl via} computer simulations.

Both the weak and the strong coupling approximations are based on a functional integral or field-theoretic representation \cite{podgornik} of the grand canonical partition function of a system composed of fixed surface charges with intervening mobile counterions, and depend on the value of a single dimensionless coupling parameter $\Xi$ \cite{Netz}.
The distance at which two unit charges interact with thermal energy $k_BT$ is known as the Bjerrum length $l_B=e_0^2/4\pi\varepsilon\varepsilon_0 k_BT$ (in water at room temperature, the value is $l_B\approx 0.7$ nm). If the charge of the counterions is $q$ then the aforementioned distance scales as $q^2 l_B$. Similarly, the distance at which a counterion interacts with the surface charge $\sigma$ with an energy equal to $k_BT$ is called Gouy-Chapman length, defined as  $\mu=e_0/2\pi ql_B\sigma$.
A competition between ion-ion and ion-surface interaction can be quantitatively measured with a ratio of both characteristic lengths $\Xi=q^2 l_B/\mu=2\pi q^3 l_B^2\sigma/e_0$, which is known as the (Netz-Moreira) coupling parameter \cite{Netz}.

The meaning of this coupling parameter can be easily understood by considering  the mean distance between counterions in the layer next to the charged surface. Assuming that electroneutrality is achieved within one Gouy-Chapman layer $\mu$, the volume available per ion is $4\pi a^3/3= qe_0\mu/\sigma$ and the mean distance between counterions $a= \mu (3{\Xi}/2)^{1/3}$. It follows that in the weak coupling case, defined by ${\Xi}\ll 1$, the width of the layer $\mu$ is much larger than the separation between two neighbouring counterions and thus the counterion layer behaves basically as a 3D gas. Each counterion in this case interacts with many others and the collective mean-field approach of the Poisson-Boltzmann type is completely justified. On the other hand in the case of the strong coupling limit, defined by ${\Xi}\gg 1$, the mean distance between counterions, $a$, is much larger than the layer width, meaning that the counterion layer behaves as a 2D gas \cite{Netz}. In this case the mean-field approach breaks down, each counterion moving almost independently from the others along the direction perpendicular to the wall and the collective effects that enable a mean-field description are thus absent. The two limits are characterised by a low/high valency of the counterions and/or a small/large value of the surface charge density. The range of validity of both limits has been explored thoroughly in \cite{Netz}.
\begin{figure}[!h]
\includegraphics[width=8cm]{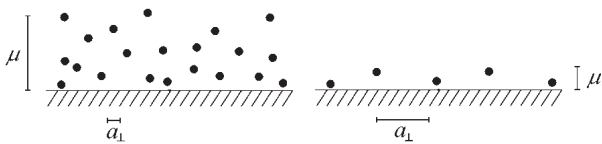}
\caption{Schematic view on the charged surface with counterions in the case of weak (lefthand drawing - $\Xi \ll 1$) and strong (righthand drawing - $\Xi \gg 1$) coupling limit. $\mu$ is the Gouy - Chapman length and $a_{\perp}$ is the average separation between counterions.}
\end{figure}

Formally the weak coupling limit can be straightforwardly identified with the saddle-point approximation of the field theoretic representation of the grand canonical partition function \cite{podgornik}, and is reduced to the PB theory in the lowest order. The strong coupling approximation has no PB-like correlates \cite{Netz} since it is formally equivalent to a single particle description, and corresponds to two lowest  order terms in the virial expansion of the grand canonical partition function. The consequences and the formalism of these two limits of the Coulomb fluid description have been explored widely and in detail (for reviews, see \cite{Naji,hoda}).

In this paper we thus embark on a study of interactions between macromolecules with pronounced helical motifs of the fixed charges that would allow us to formulate their interactions on the strict level of weak and strong coupling approximations. Our goal here is not to go into detailed modeling of the charge distribution as was done in the case of DNA by Kornyshev and Leikin \cite{Kornyshev}, neither do we intend to study the details of ss-DNA wrapping geometry and energetics in the case of decorated CNTs as was pursued by Lustig {\em et al.} \cite{JagotaLustig}. Our goal is more modest: we intend to asses the consequences of the weak and strong coupling dichotomy as it transpires through the interactions between macromolecules with a helical charge motif in the presence of counterions, and consequently to explore the specific features the helical charge motif adds to the interactions in both limits, studied extensively heretofore \cite{hoda}. A similar, yet in fundamental respect different point of view, was recently advanced by Lee \cite{dominiclee}.

\section{The model}
We consider a model of two identical infinite parallel charged cylinders with radius $a$ and interaxial separation $R$. The charge on both cylinders is distributed on a single helix with a pitch $H$. In order to avoid spurious divergences when solving the PB equation, we represent this helical charge distribution as a helical stripe with small but finite thickness. This can be justified by the finite size of the fixed charge groups on the macromolecular surface, as well as the finite size of the counterions.

Both cylinders can be rotated around their axes by angles $\varphi_1$ and $\varphi_2$. Without loss of generality we will put $\varphi_1\equiv 0$ and $\varphi_2\equiv \varphi_0$, which just shifts the origin of the coordinate-system. For continuous helices the rotation of the second cylinder by an angle $\varphi_0$ corresponds to a translation by $\Delta z=(H/2\pi)\varphi_0$ along the direction of the axis of the cylinder.
\begin{figure}[!ht]
\centerline{\psfig{figure=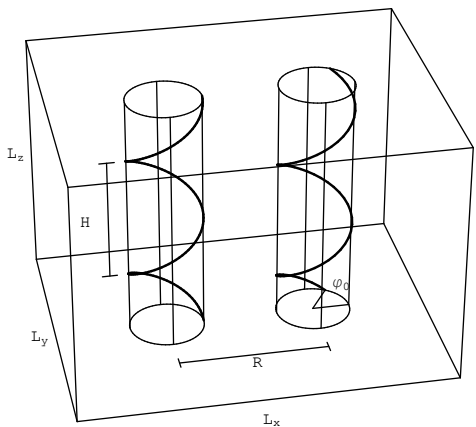,width=6.5cm}}
\caption{Two cylinders with (single) helical charge distribution in a bath of counterions (not shown).}
\label{skica}
\end{figure}
The charge of both cylinders is compensated by freely mobile counterions with valency $q$ bathing the cylinders. In counterion-only case we neglect all coions. This approximation is relevant for low salt concentrations where the Debye screening length is much larger than the scales of interest. We consider the cylinders as impenetrable to counterions. In our model the cylinders are considered as hollow, so that they contain water and therefore have the same dielectric constant as the bulk water, $\varepsilon=80$. This can be justified for hollow CNTs, whereas it is only a rough approximation for a DNA molecule, because it neglects image contributions \cite{Kornyshev}. However, even if we  include exact image effects this would still lead to a very rough approximation of a real DNA molecule because of unknown surface and saturation effects on the value of the local dielectric constant. We will not delve into these complicated and poorly understood effects in this work.

The amount of charge on both cylinders can be expressed with a dimensionless Manning parameter \cite{najicylinder, Trizac}, defined as
\begin{equation}
Q=\frac{q^2 l_B}{d}
\end{equation}
where $d$ is the longitudinal spacing between equivalent elementary charges along the cylinder and $l_B$ is Bjerrum length.
Expressing the Manning parameter with mean surface charge density $\sigma$, one remains with
\begin{equation}
Q=\frac{e_0qa}{2\varepsilon\varepsilon_0 kT}\sigma.
\end{equation}
Note that at fixed surface charge the Manning parameter $Q$ is proportional to counterion valency $q$.

We will solve the statistical mechanics of the above model in two well defined limits \cite{Naji}: the weak-coupling (WC) a.k.a. the Poisson-Boltzmann (PB) or the mean-field limit and the strong-coupling (SC) limit. Their pertinent range of validity have been analysed thoroughly, see {\sl e.g.} \cite{Naji}. These two coupling regimes delimit the exact thermodynamic properties of the system and thus describe the extreme limits of its behavior in the parameter space. We will not include in our analysis the contribution of the second order fluctuations around the mean-field solution \cite{rudi-first}, as can be done for planar systems, since it is a lot more complicated in cylindrical geometry \cite{rudi-second}, while remaining numerically overall small.

\section{Weak coupling limit (mean-field)}
The mean-field approach, also known as weak-coupling limit ($\Xi\ll 1$), is based on the Poisson-Boltzmann equation for charged mobile counterions in solution \cite{Andelman} and is valid for $\Xi\ll 1$. This theory predicts that counterions will distribute in the space surrounding the macromolecular charge in accordance with the Boltzmann statistics that leads to the following equation for the mean electrostatic potential
\begin{equation}
\nabla^2\phi=-\frac{A }{\varepsilon\varepsilon_0}\,e^{-\beta e_0 q\phi},
\end{equation}
where the constant $A$ can be determined by the charge neutrality condition. Inside the cylinders the right-hand side of the above equation is $0$. Using the dimensionless electrostatic potential, $u=\beta e_0 q \phi$, we obtain a set of equations valid inside and outside the cylinders
\begin{equation}
	\begin{array}{ll}
	\nabla^2 u=-C e^{-u}&\textrm{(outside),}\\
	\nabla^2 u=0&\textrm{(inside)}.
	\end{array}
	\label{prva}
\end{equation}
The boundary conditions on the surface in the presence of the surface charge density $\sigma^*$, that can exhibit spatial variation along the surface,  are formulated in the standard way {\sl via} the normal component of the electric field strengths as $$\varepsilon\varepsilon_0 E_n-\varepsilon\varepsilon_0 E'_n=\sigma^*,$$where $\varepsilon$ is the (static) dielectric constants outside as well as inside the cylinders. In dimensionless variables this amounts to
\begin{equation}
	\frac{\partial u'}{\partial r}\Big\vert_\partial-
	\frac{\partial u}{\partial r}\Big\vert_\partial=
	\left\{
	\begin{array}{cl}
		\cfrac{\beta e_0 q}{\varepsilon\varepsilon_0}\,\sigma^*,&\textrm{charged stripe},\\
		0,	&\textrm{otherwise.}
	\end{array}
	\right.
	\label{druga}
\end{equation}
In order to avoid numerical divergences in the derivatives if the surface charge density $\sigma^*$ is represented by an infinitely thin line, we represent the helical charge distribution as a helical stripe with a finite thickness. As already noted this can be justified by realizing that a typical ion radius would be around $0.2$ nm and therefore we assume the stripe of thickness $0.4$ nm.

Numerically we solve the PB equation in a finite bounding box geometry with dimensions $L_x\times L_y \times L_z$ (see ~\ref{skica}). The height of the box is always taken to be equal to one pitch,$L_z=H$ and the periodic boundary conditions are applied along the $z$ direction to mimic infinitely long cylinders. Results depend on lateral box sizes $L_x$ and $L_y$ and should converge when these two sizes are large enough.

Solving for the dimensionless potential $u$ we can calculate the force acting between both cylinders {\sl via} the stress tensor $p$, composed of the Maxwell part and the ideal gas (van't Hoff) part as
\begin{equation}
p_{ik}=\varepsilon\varepsilon_0\left(E_iE_k-\frac{1}{2}E^2\delta_{ik}\right)-p\,\delta_{ik}.
\end{equation}
The van't Hoff pressure of the counterions, $p$, equals their ideal gas pressure. The force between cylinders can be evaluated by integrating the stress tensor over any closed surface embedding one of the cylinders:
\begin{equation}
F_i=\oint_S p_{ik} dS_k \;.
\end{equation}
We choose to integrate over the mid-plane between the cylinders and over the outer edges of the box, as was done in \cite{akesson}. With large enough simulation box the integral over the outer edge should become negligible.

As stressed already in the introduction the helical charge model can be considered as an approximate rendition of either the ds-DNA or the ss-DNA/CNT hybrid system. For the ds-DNA the structural parameters are: cylinder diameter of $a=1~$nm and helical pitch of $H=3.4~$nm,  so that the Manning parameter is $Q=4.1q$ where $q$ is the counterion valency. In the case of ss-DNA/CNT hybrid complex, the cylinder radius can typically be in the range of $1.5-3~$nm and the pitch from about $3.4~$nm on \cite{JagotaManohar,JagotaLustig}. Depending on these numerical values, the Manning parameter in the ss-DNA/CNT case is in the range of $2-10$ for monovalent counterions. If not stated otherwise the counterion radius is taken to be $R_c=0.2~$nm.
Since ds-DNA case is in the middle of both ss-DNA/CNT extremes we will mostly delimit ourselves to DNA parameters as typical values used in the numerical calculations. They are used for illustrative purposes and not for an explicit description of ds-DNA {\sl per se}.

The coupling parameter for the ds-DNA case is $\Xi=2.8~q^3$ in the presence of counterions with valency $q$. The results for WC as well as SC approximations do not depend directly on $\Xi$ but this parameter determines the range of validity of these approximate theories \cite{Naji}. For monovalent counterions the coupling parameter has a value of $\Xi=2.8$ which is small enough to justify the WC approach \cite{Chan}. But this approach certainly breaks down for tri- and tetra-valent case where $\Xi=76$ and $180$, respectively.

\subsection{Results}

In order to solve numerically the system of partial differential equations with spatially varying boundary conditions, Eqs. \ref{prva} and \ref{druga}, we used the COMSOL Multiphysics package, a finite element analysis and solver software, together with the  Cholesky decomposition method for evaluating results on finite element mesh.

Before we delve into helical charges we take a quick glance at the behaviour of two homogeneously charged cylinders \cite{najicylinder}. First of all we want to explore the effect of the finite bounding box in our numerical calculations. The bounding box has a square base, $L_x=L_y=L$ and height $L_z=H$ with periodic boundary conditions applied in the vertical direction. It is well known, that the counterion density around a single uniformly charged cylinder behaves asymptotically as $r^{-2}[1+(Q-1)\ln\,r/a]^{-2}$ \cite{Lifson,Hansen1}. Therefore, we expect to see the same asymptotic behaviour for helical charge distribution at large distances also in our numerical calculations, with the proviso that $Q \to 2Q$ for two cylinders.

\begin{figure}[!ht]
\centerline{\psfig{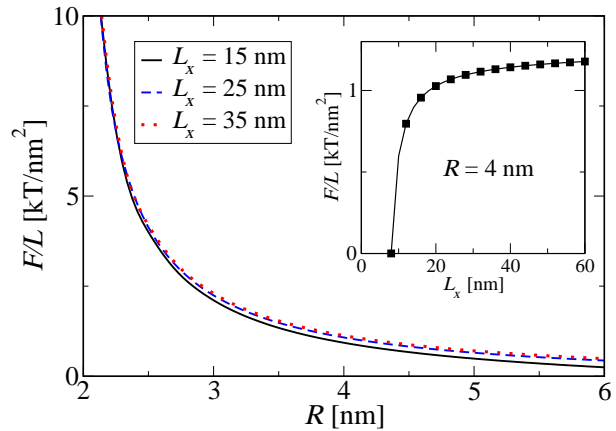}}
\caption{Force per unit length - separation dependence for uniformly charged cylinders with ds-DNA parameters ($q=1$) for different box sizes $L_x=L_y=L$. The finite box size effect is of course more pronounced at larger separations. The inset directly shows the variation of the force with increasing box size at the distance $R=4~$nm. One should note the logarithmically slow convergence of the result.}
\label{WCForce2D}
\end{figure}

The density profile in our case decays very slowly, causing weak convergence with increasing box size. For this reason the finite-bounding box effects are important. This size dependence can be discerned from Fig~\ref{WCForce2D}. By constraining the space around the macroions we artificially heap surrounding counterions into the box that should otherwise extend further away.
At small interaxial separations $R$ the force is obviously not as sensitive to the bounding box size as at larger separations. As shown in the inset of Fig~\ref{WCForce2D}, the convergence of the result is very weak when $L_x$ exceeds  $\approx {\rm 20} ~a$. In all of the calculations presented below we used $L= {\rm 25}~ a$ and we have to be aware that this result is slightly but not fundamentally different from the strict formal limit  of an infinite box size.

On the mean-field level the force between charged helices is always repulsive and decays with increasing interaxial distance.
In a symmetric system on the mean-field level it is well known that interactions are always repulsive, a result that can be formulated as a general theorem \cite{Chan}.

\begin{figure}[!ht]
\centerline{\psfig{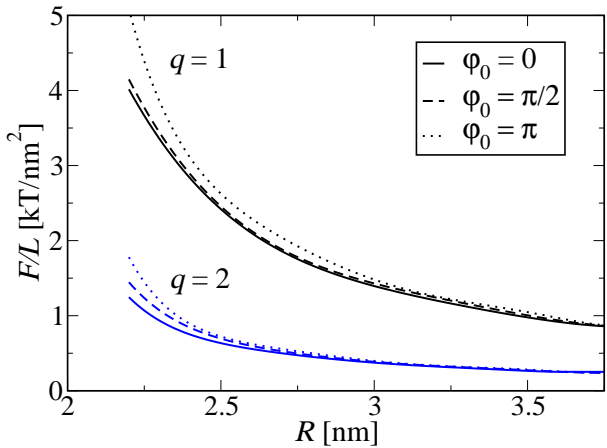}}
\caption{Force per unit length - separation dependence for two helices (ds-DNA parameters) at various mutual azimuthal angles $\varphi_0$ and valencies. The corresponding Manning parameter for $q=1$ is $Q=4.1$ and for $q=2$ it is $Q=8.2$.}
\label{WCForceR}
\end{figure}

\begin{figure}[!ht]
\centerline{\psfig{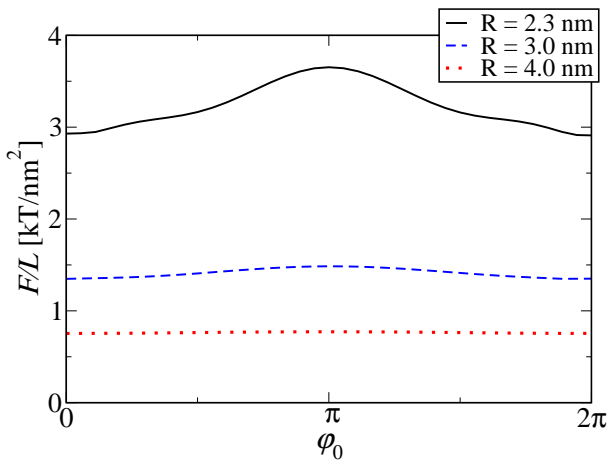}}
\caption{Azimuthal angle dependence of the force per unit length at different interaxial distances for ds-DNA parameters and monovalent counterions ($Q=4.1$).}
\label{WC-phi}
\end{figure}

On Fig~\ref{WCForceR} and \ref{WC-phi} we present the results of numerical calculations of the interaction force between two cylinders with a helical charge motif as a function of the interaxial separation as well as the azimuthal angle of mutual orientation. Increasing the Manning parameter (corresponding to higher counterion valencies) decreases repulsive interaction but  can not lead to a change of sign of the interaction. The maximal angular variation at a separation of  $R=2.3~$nm amounts to only $16 \%$ of the total force per unit length, and is a lot smaller if the cylinders are even further apart.

The azimuthal angle dependence of the force for two helical charge distributions, Fig~\ref{WC-phi}, thus shows only a slight variation, being larger at smaller interaxial separations. This is a completely general conclusion of the WC analysis. The azimuthal modulation of the force has a maximum at a configuration corresponding to $\varphi_0=\pi$ and a minimum at $\varphi_0=0$. It also shows mirror symmetry along $\varphi_0=\pi$ line, so that configurations $\varphi_0$ and $2\pi-\varphi_0$ have exactly the same interaction.

\section{Strong coupling limit}
In the limit of $\Xi\gg 1$ the mean-field {\sl ansatz} breaks down and a different kind of approach is needed. Here, virial expansion of the partition function in terms of fugacity $\lambda'$ can be used \cite{Netz}, leading to the strong-coupling limit. For our purposes only the expansion of partition function ${\cal Z}_G$ to the first order will be used, that effectively amounts to
\begin{equation}
{\cal Z}_G={\cal Z}_G^{(0)}+\lambda'{\cal Z}_G^{(1)}+{\cal O}(\lambda'^2).
\end{equation}
Here the zeroth-order term corresponds to the bare electrostatic interaction energy of charged cylinders, and the first-order term corresponds to the one particle counterion contribution to the partition function. For details of the strong-coupling approach, see \cite{Naji} and references therein.

To obtain the free energy in the SC limit we follow the same procedure as in \cite{Kanduc}. The first step is to evaluate the grand potential from the grand partition function and then by using a Legendre transformation we get the free energy per counterion  as
\begin{equation}
\frac{\beta {\cal F}}{N}=\frac{\beta W_0}{N}-\ln\int_V e^{-\beta u(\Av r)}dV.
\label{SCFree}
\end{equation}
The first term in the above expression, $W_0$, represents the bare electrostatic energy between charged helices.
It can be evaluated simply by the integration of charge contributions on both helices with the unscreened Coulomb kernel
\begin{equation}
\frac{\beta W_0}{N}=\frac{Q}{2H}\int_{-\infty}^{\infty}\int_0^H\frac{dz\,dz'}{s(z,z')}.
\end{equation}
Here, $s(z,z')$ represents the distance between two points on different helices with coordinates $z$ and $z'$, respectively.
Helices are represented as (infinitely) thin lines and not as a stripes, as was the case in the WC case. On the SC level the counterion hard core radius will be taken into account {\sl via} its excluded volume in the partition function \cite{Kanduc}. The second term in Eq. \ref{SCFree} is a one-particle contribution to the free energy. Here $\beta u$ is the electrostatic energy of a single counterion with charge $e_0 q$ in the presence of two charged helices. It is obtained simply by integration of the Coulomb kernel along both helices as
\begin{equation}
\beta u=-Q\int_{-\infty}^{\infty}\left(\frac{1}{s_1(z')}+\frac{1}{s_2(z')}\right)dz',
\label{u}
\end{equation}
where $s_1$ and $s_2$ are distances from the counterion-coordinate to the point on helix $1$ and helix $2$ defined as
\begin{eqnarray}
&&s_{1,2}(\rho,\varphi,z;z')=\\\nonumber
&&\hspace{3ex}\sqrt{a^2+\rho^2-2a\rho\,\cos(kz'+\varphi_{1,2}-\varphi)+(z-z')^2}\
\label{eq-s}
\end{eqnarray}
The angles $\varphi_1$ and $\varphi_2$ represent orientations of both cylinders around their symmetry axes. Here, again as in PB case, the convention $\varphi_1=0$ and $\varphi_2=\varphi_0$ is used without loss of generality.

The integral in Eq. \ref{SCFree} should be in principle evaluated over the entire space available to the counterions, but we again delimit integration inside the confining box. The height of the box is again taken as one helical pitch, $L_z=H$, due to the periodicity of the potential $u(\Av r)$ in $z$-direction. Any multiple value of $H$ in $L_z$ would produce only an additive constant in the final free energy result. The problem of finite integration box in the strong-coupling limit is not so critical as in the Poisson-Boltzmann case, since counterion density decays very rapidly with radial distance. In the case of a single uniformly charged cylinder the density decays as $r^{-2Q}$, so the asymptotic behaviour at large distances for two cylinders with a helical charge motif behaves as $r^{-4Q}$. Typically, $Q>4$ for DNA so the volume integral converges very rapidly. A rectangular box of sides $L_x=7a$ and $L_y=3a$ was used, which is large enough to produce results very close to the infinite box size limit.

Hard core repulsion of cylinders as well as finite counterion-size are taken into account {\sl via} the hard-core radius of the cylinders, equal to $a+R_c$. Here $R_c$ is counterion radius and $a$ the bare radius of the cylinder.

The interaction force between both cylinders is obtained by taking the derivative of the free energy with respect to the interaxial separation
\begin{equation}
F=-\frac{\partial {\cal F}}{\partial R}.
\end{equation}
Obviously,  the force scales with the length of the cylinders, therefore we express the force per length of cylinders, $F/L$.

\subsection{Results}
\begin{figure}[!h]
\includegraphics[width=8cm]{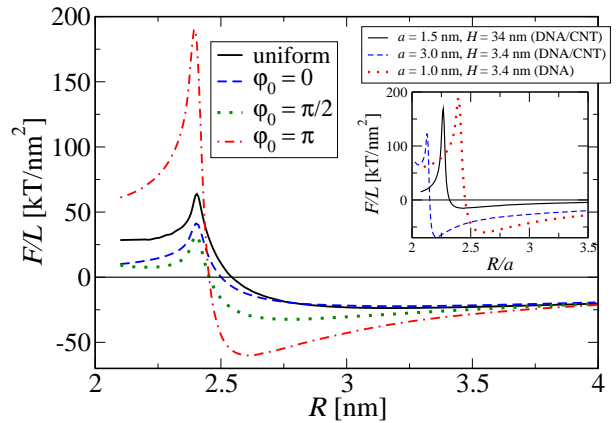}
\caption{SC force per unit length - separation curves for various orientations $\varphi_0$ in the case of ds-DNA parameters of $a=1$ nm and $H=3.4$ nm in the presence of monovalent counterions. The corresponding Manning parameter is $Q=4.1$ and the coupling parameter $\Xi\simeq 3$. The inset shows force per unit length - separation curves for orientation $\varphi_0=\pi$ for different cylinder radii, corresponding to the numerical values of the parameters of the ss-DNA/CNT hybrids. Qualitatively the interactions are the same.}
\label{SCForce}
\end{figure}

The interaction force in the case of the strong-coupling limit shows a much richer behavior than in the weak-coupling (PB)  case.
As seen from Fig~\ref{SCForce} the interaction force for various parameters, including the ds-DNA as well as ss-DNA/CNT values, exhibits similar qualitative behaviour. Again we will use ds-DNA parameters for illustrating our analysis. We present the SC results for monovalent as well as polyvalent counterions even though small valencies $q$ that correspond to small coupling parameter $\Xi$ are not relevant for the SC theory. The aim here is to illustrate different behaviour of the interaction for counterions of different valencies, {\sl i.e.} of different  Manning parameters.

We find three different regimes in the behavior of the  interaction as a function of interaxial separation, Fig~\ref{SCForce}. At large separations the SC force is always attractive. This is due to the fact that each counterion is localized at one or the other cylinder, effectively creating a correlation hole, causing net correlation attraction between the helices. This is indeed very similar to the interaction of planar surfaces \cite{Naji}. At larger interaxial spacings the SC theory breaks down and the WC repulsion takes over.

Similar to the case of interacting planar surfaces, for very small interaxial spacings the counterions accumulate in the space between the two cylinders, creating a repulsive component to the total interaction due to their osmotic pressure. The osmotic contribution can be larger than the electrostatic correlation attraction and a net repulsion ensues for the total interaction between helices at very small separations.

\begin{figure}[!h]
\includegraphics[width=8cm]{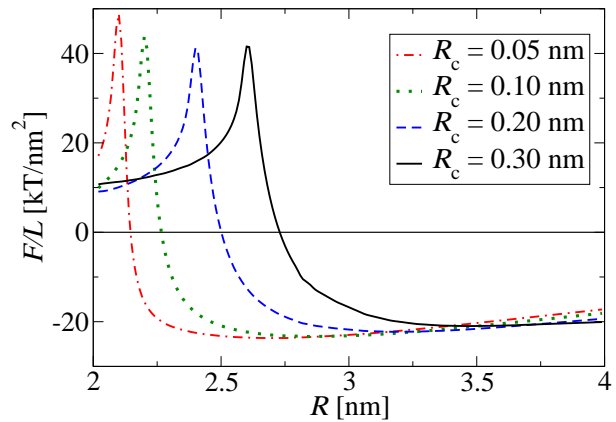}
\caption{The hard core counterion-radius effect. Depletion region, where the force is reduced, appears at distance smaller than counterion diameter.}
\label{ForceRc}
\end{figure}

The highest value of osmotic pressure is reached when the separation between cylinders becomes equal to the counterion diameter $2R_c$, see Fig~\ref{ForceRc}. This leads consequently to the highest repulsive force between the cylinders. At yet smaller separations, the depletion effect sets in, in the sense that the counterions can not penetrate anymore the inter- cylinder space, and are thus depleted from the spatial region between the two apposed cylinders, diminishing their osmotic pressure in that region. Thus, the counterion contribution to the osmotic pressure is reduced, {\sl i.e.} the osmotic repulsion as well as counterion electrostatic interactions are both reduced. In all these situations the bare electrostatic repulsion between the two helices remains unaffected. Because of this, at a separation $R^*$ where the net force is zero $F=0$, the bound state between the cylinders occurs.

The relative orientation of the two cylinders  plays a crucial role in determining the nature of the inter-helical interaction. The force at orientation $\varphi_0=\pi$ now appears an order of magnitude larger than at orientations $0$ or $\pi/2$. This is quite different than in the WC case, where the analogous variation was much smaller, and indeed almost negligible. The case of two uniformly charged cylinders is found somewhere in between both extremal behaviors, see Fig~\ref{SCForce}.

\begin{figure}[h]
\includegraphics[width=8cm]{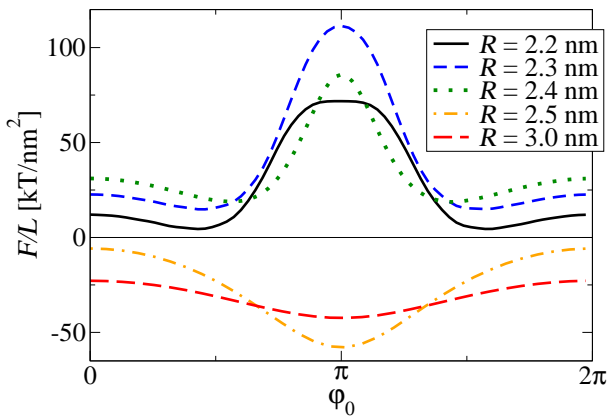}
\caption{Force per unit length - angle curves for various interaxial distances. DNA parameters for $q=1$ are taken.}
\label{SC-Fphi}
\end{figure}

From Fig~\ref{SC-Fphi} we also deduce that the extremum of the force is reached at $\varphi_0=\pi$. This could be explained by the fact that at $\varphi_0=\pi$ the charges on helices are apposed and can approach at a smallest local separation.
The electrostatic potential between the cylinders is therefore higher than for other values of $\varphi_0$, which causes stronger localization of counterions in between. Stronger localization in its turn leads to larger attraction at large distances as well as greater osmotic repulsion at smaller distances.

As seen from Fig~\ref{SC-q} the counterion valency, {\sl i.e.} Manning parameter $Q$, can also have quite a dramatic effect in the SC limit. Note that increasing the valency $q$ also increases the Manning parameter in the same proportion.
At large distances the force is independent of the valency. This makes sense, since at large distances each counterion is located either at one or the other cylinder. In this case counterions cannot sterically interact with both cylinders, therefore the osmotic repulsive contribution vanishes. Only electrostatic interaction thus remains that depends on the net counterion charge but not on the amount of counterions, {\sl i.e.} it does not depend on counterion valency. Note, that higher valency means smaller amount of counterions since the net counterion charge remains the same due to electroneutrality.

The influence of counterion valency is significant also at smaller interaxial separations where osmotic component plays an important role. Each counterion contributes to the osmotic pressure in the vicinity of cylinders. Its  contribution is larger for higher electrostatic potential around cylinders as well as larger valency. Namely, counterions with larger valencies are more attracted and localized in the vicinity of cylinders. This dependence of the osmotic force on the valency is non-trivially connected with the electrostatic potential. The net osmotic force in SC is the product of single-counterion contribution and the amount of counterions $N$. Due to electroneutrality condition, this amount is then inversely proportional to the valency, {\sl i.e.} $N\propto q^{-1}$.
The net osmotic contribution can thus be an increasing as well as a decreasing function of $q$.

\begin{widetext}\phantom{}
\begin{figure}[t!]
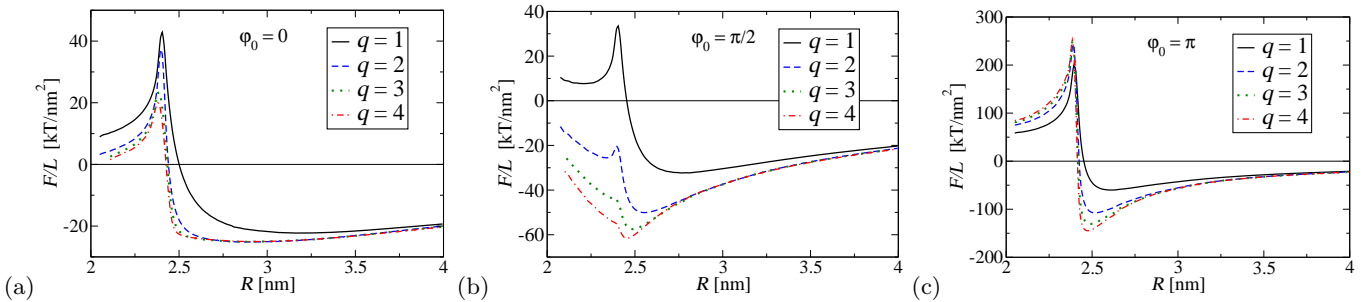

\centerline{
(a)\includegraphics[width=5.5cm]{Forceq1.eps}
(b)\includegraphics[width=5.5cm]{Forceq2.eps}
(c)\includegraphics[width=5.5cm]{Forceq3.eps}
}
\caption{Force-separation curves for various counterion valencies $q$ at different orientations $\varphi_0$ (ds-DNA parameters). At large distances the force is independent of the valency.}
\label{SC-q}
\end{figure}
\end{widetext}

Fig~\ref{SC-q} shows the dependence of the force per unit length on the interaxial spacing for various orientations as well as various counterion valencies. In (a) the repulsive osmotic contribution at $\varphi_0=0$ is apparently higher for smaller valencies and its value per counterion increases slower than $q$.
We found out, that such a trend also appears if the cylinders are uniformly charged, the case analyzed in \cite{najicylinder}.
In the $\varphi_0=\pi$ case, the opposite is true  (Fig~\ref{SC-q} (c)). In this case, the segments of charged helices approach to the smallest mutual separation, causing much higher electrostatic potential in between, that enhances counterion localization  which furthermore leads to higher osmotic repulsion for larger valencies.
Thus on increasing $q$ the enhanced localization leads to more pronounced growth of the osmotic contribution per counterion so that the net osmotic force increases with increasing $q$. In both configurations $\varphi_0=0$ and $\varphi_0=\pi$ the largest potential appears exactly in the middle of both cylinders which induces a maximal counterion density at the same position. This is however not the case for other orientations.

\begin{figure}
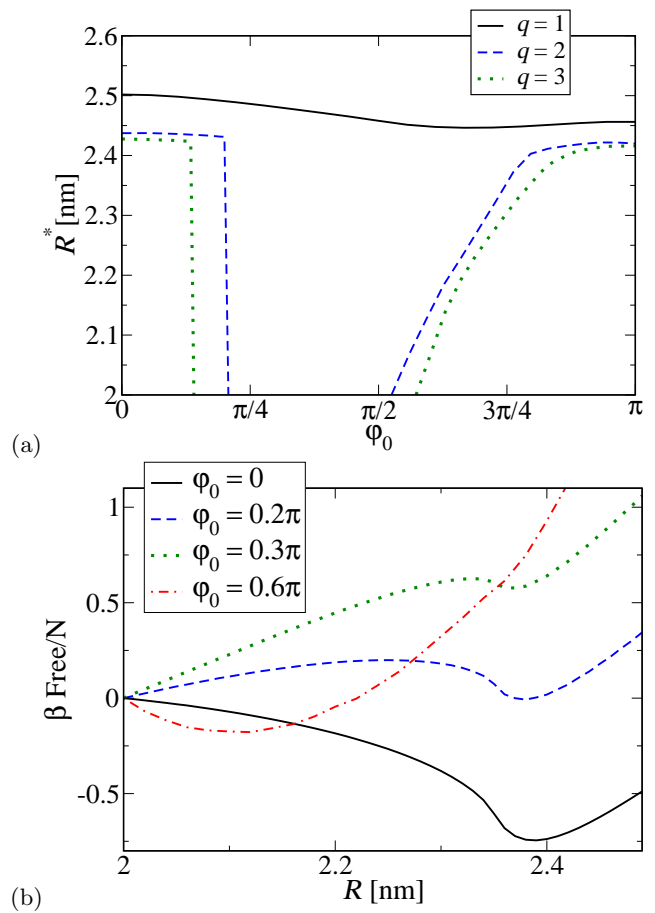

(a)\includegraphics[width=8cm]{Rbound.eps}
(b)\includegraphics[width=8cm]{FreeEn.eps}
\caption{Collapse transition into a bound state for ds-DNA parameters. (a) Bound state interaxial separation as a function of mutual orientation $\varphi_0$. For polyvalent ions the rapid jump appears at certain angle, corresponding to phase transition. (b) Corresponding free energy-distance plot for divalent counterions ($q=2$) at different orientations. Two minima (for the first three cases) are clearly visible which are responsible for the phase transition.}
\label{Rbound}
\end{figure}

At the configuration $\varphi_0=\pi/2$ (Fig~\ref{SC-q} (b)) an even more dramatic phenomenon occurs. Here, the highest potential appears slightly out of the interaxial plane. The counterions that are localized slightly away from the central region between the helices also contribute less to the osmotic pressure. For $q=1$ ($Q=4.1$) case when the amount of counterions is still large enough and localization not so strong the osmotic pressure prevails over the electrostatic attraction at smaller separations. But for $q\ge 2$ ($Q>8$) the attractive electrostatic interactions prevail at all separations, leading to a collapse transition of both cylinders into a bound state, corresponding to the closest approach distance, which is the double radius of the cylinders $2a$.  As seen from Fig~\ref{Rbound}a, the bound-state distance $R^*$ changes only slightly for small Manning parameters, corresponding to monovalent counterions for the ds-DNA case.
The occurrence of this transition can be understood from the free energy-interaxial distance plot in Fig~\ref{Rbound} (b). The free energy exhibits two minima at small orientations $\varphi_0$, with the minimum at finite separation being the global one. On increasing $\varphi_0$ the free energy minimum located at the closest separation $R=2a$ becomes deeper, leading to a discontinuous transition into a bound state. At even larger orientations the minimum continuously moves away from the closest separation to finite separation.

Around $\varphi_0=0$ and $\varphi_0=\pi$ the bound-state distance is very close to the case of two uniformly charged cylinders. There, a simple analytical expression can be derived, $R^*-2a=2R_c+\frac{2}{3}\mu$ \cite{Naji2}. The difference in separations between helical and uniform charge distribution is within $20\%$ for monovalent and less than $10\%$ for polyvalent case for aforementioned orientations. But it completely fails for intermediate $\varphi_0$ for polyvalent case.

We also want to briefly comment on the behaviour for small Manning parameters. According to Naji {\em et al.} \cite{Naji2}, a significant dependence upon the size of the confining box appears for Manning parameters smaller than $Q_c=2/3$ in the case of two uniformly charged cylinders. This occurs as a result of the dilution of the counterion cloud around weakly-charged cylinders. Below the critical value of the Manning parameter, $Q_c$, the equilibrium distance scales with the box size approximately as $R\sim L_x/\sqrt{\pi}$. We found that helical charge distribution in our case exhibits the same behaviour below the critical Manning parameter $Q_c=2/3$. This is expected, since the charge distribution details are not important at large separations. Such small values of $Q$ do not of course correspond to DNA parameters.

Within our model there is no attraction between the helices on the mean-field level, so there can not be any bound state.
In the ds-DNA model of Kornyshev and Leikin \cite{Kornyshev, KornyshevPRL} the counterions are assumed to be localized and fixed on the surface of the helix, leading to a charge separation between negative charges of the phosphates and positive charges of the (fixed) counterions. In that case of course, even on the linearized PB level, one can still see attractions due to charge separation, leading to a bound state with a typical equilibrium separation of 0.5 nm, depending on the parameters - close to the experimentally reported value in bundles of DNA molecules \cite{Rau}. Fixing the counterions on the DNA surface of course assumes an additional specific binding interaction, not present in our model where the only interaction is Coulombic, so the results for the Kornyshev-Leikin and our model can not be compared directly at this point.

If the helices are rotationally unconstrained, they will always settle into an orientation that corresponds to a minimum of their free energy. In the weak-coupling case this orientation always corresponds to $\varphi_0=0$, since in this configuration the helical segments are locally maximally apart and the (repulsive) free energy is thus minimal. This was already emphasized by Kornyshev and Leikin \cite{Kornyshev, Harreis} in the linearized PB approach for two single helices.

However, the behavior of this optimal azimuthal angle in the SC case is very different and in many respects opposite to that observed on the WC linearized PB level \cite{Kornyshev, Harreis}. At large distances the optimal angle is $\varphi_0=\pi$. This is in fact the most {\em unfavorable} configuration in the WC case. Correlation effects in SC dramatically change the behaviour of the system. Here, counterions mediate electrostatic correlation  attraction between both helices and thus they tend to orient themselves so that the local helical segments approach as much as possible. This happens exactly at the $\varphi_0=\pi$ configuration. We must emphasize here that only the electrostatic (correlation) interaction plays a role here while the repulsive osmotic contribution to angular interaction is negligible since it cannot cause any torque, acting always perpendicular to the cylindrical surface. That is the reason that the optimal angle remains $\varphi_0=\pi$ even though the net force turns repulsive!

\begin{figure}[!h]
\includegraphics[width=8cm]{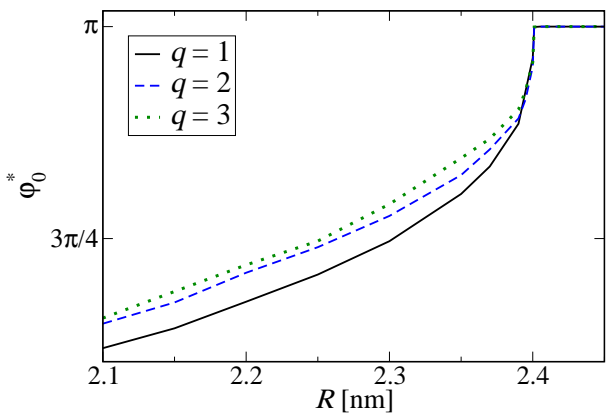}
\caption{The interaxial distance dependence of the optimal angle between two helices with DNA parameters. The critical separation where the rapid change of optimal angle occurs equals the counterion diameter, $R-2a=2R_c=0.4~$nm.}
\label{opt-phi}
\end{figure}

Everything said heretofore remains valid for larger surface-to-surface cylindrical separations, when compared to the counterion diameter $2R_c$. At smaller separations the counterions are depleted away from the mid-region between the cylinders. The localization of counterions is then out-of-center and causes a torque on helices resulting in a different value of the optimal angle $\varphi_0^*$. As shown on Fig~\ref{opt-phi} the optimal angle above the critical separation ($2R_c-2a=0.4~$nm) is $\pi$. Below this separation a substantial variation of the angle can be discerned. Note, that configurations $\varphi_0$ and $2\pi-\varphi_0$ are equivalent, so we always plot the one with the smaller azimuthal angle.

The reason for the variation of the optimal angle in SC at smaller separations is the depletion effect that is not present in the mean-field case. Similar behaviour of the optimal angle below a critical value of the interaxial separation was also found on the linearized PB level for multistranded helices. There, the spontaneous symmetry break occurs at a critical separation where the optimal angle switches from $\varphi_0=0$  to higher values \cite{Kornyshev}.

\subsection{Regime of applicability of SC results}
So far, we evaluated numerical results for the strong coupling limit for charged helices without any justification of its validity. In a planar system with a finite value of the coupling parameter $\Xi$, the asymptotic strong-coupling results hold exactly as long as the surface separation $\delta$ is smaller than the typical lateral distance between counterions $a_{\perp}$ \cite{Naji}.
This condition in fact yields a simple and generic criterion identifying the regime where strong-coupling attraction is expected to emerge between two like-charge planar macroions. It was originally suggested by Rouzina and Bloomfield \cite{Rouzina}.
In the case of uniformly charged cylinder with a  large coupling parameter $\Xi$, counterions tend to line up on opposing surfaces of the cylinders and along axes forming a correlated inter-locking pattern \cite{Deserno}. Typical distance between counterions along the axis of the cylinder, $a_z$, may be estimated from the electroneutrality condition
\begin{equation}
a_z=\frac{e_0q}{2\lambda}.
\end{equation}
The strong coupling limit is expected to become valid when the surface-to-surface distance of the cylinders, $\delta=R-2a$, becomes smaller than the distance between counterions, {\sl i.e.} $\delta<a_z$, leading to the criterion \cite{AliArnold}
\begin{equation}
\delta<\frac{\Xi}{Q}~\mu=\frac{l_B}{2Q_1}~q,
\end{equation}
where $Q_1$ is the Manning parameter for monovalent counterions. Note that this critical separation grows linearly with valency $q$. Taking the parameters for ds-DNA, with $Q_1=4.1$ and Bjerrum length $l_B=0.7~$nm the values of critical distance $\delta$ are in the range from 0.2 nm to 0.7 nm when the valency $q$ goes from from 1 to 4. This calculation suggests that tri- and tetra-valent counterions do justify the SC approach at separations around 0.5 nm and below, where most of the interesting phenomena described above occur anyhow. But according to this criterion, monovalent and divalent counterions can not be described within the SC limit for the ds-DNA parameter case.

\begin{figure}[!ht]
\centerline{\psfig{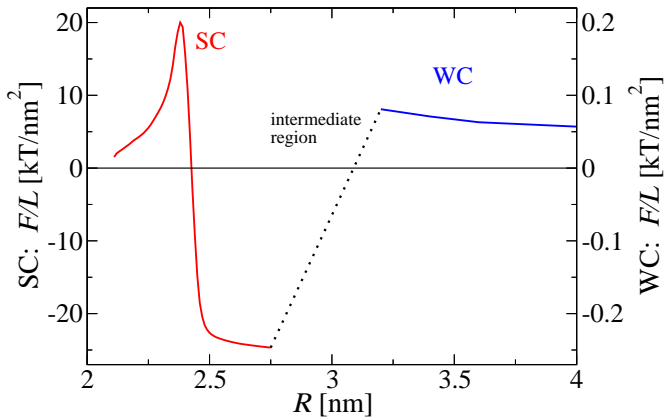}}
\caption{Schematic representation of SC and WC regions for ds-DNA parameters. $\varphi_0=0$ in the presence of tetravalent counterions, $q=4$. In the intermediate separations no approach is valid and one would have to take recourse to extensive simulations.}
\label{SCWC}
\end{figure}

Note that we used the criterion for uniformly charged cylinders and that helical charge pattern might change this criterion since counterions are localized and line up differently. The regime of SC-validity might also depend on orientation of molecules. But so far we must rely on this simple and rough criterion.

\section{Conclusion}

In this paper we have analysed the electrostatic interactions between two cylinders with a helical charge distribution in the presence of counterions. We used the approach {\sl \' a la} Netz and coworkers by explicitly considering the limits of weak and strong coupling, defined by the value of the electrostatic coupling parameter $\Xi$, which is assumed to be small in the former and large in the latter case.

The weak coupling case, or equivalently the mean-field case, is relevant when counterion valency is small, {\sl i.e.} $q=1$, or at large inter-helical separation. The distribution of counterions and the corresponding electrostatic potential is governed by the Poisson-Boltzmann equation. The force between two charged helices has been evaluated by the stress tensor method and appears to be overall repulsive in the WC case, a result known from previous works \cite{Oosawa, Ohnishi, Chan}. Helical charge distribution here makes up  for only a slight angle modulation with comparison to the uniformly charged case. The largest repulsive force appears at $\varphi_0=\pi$ mutual cylindrical orientation and the smallest at $\varphi_0=0$. Increasing the counterion valencies $q$ tends to decrease the amount of counterions in the system and the corresponding osmotic contribution as well as the net interaction is therefore smaller but it never changes the sign.

In contrast to weak coupling, the strong coupling limit exhibits a much richer and a lot more interesting behavior. The SC theory  is effectively a one-particle theory \cite{Naji} that neglects all interactions among counterions and only takes the interactions between the counterions and the fixed charges on the macromolecular surface, {\sl e.g.} the ds-DNA surface, into account. This approach is justified if the distances between counterions are much larger than their distances between the cylinders. In other words, SC limit is valid for large counterion valencies $q$ and for small separations between the cylinders. Particularly for ds-DNA parameters, the SC approach for $q=3$ and $4$ would be valid for surface-to-surface separations under $\sim 0.5~$nm which is comparable to typical counterion diameters.

As we have been arguing above, the SC force between two cylinders is always attractive at larger separations, but can become repulsive at smaller separations due to the osmotic counterion contribution. At separations smaller than counterion diameter the counterions are depleted away from inter-helical region which reduces the osmotic repulsion. At the separation where the net force is zero, a bound state occurs. This bound-state separation approximately equals to the counterion diameter, typically $0.4~$nm.

On the SC level the relative orientation of cylinders has a drastic effect on interactions between cylinders with a helical charge distribution. In many respects the interactions between helices in this limit are opposite to those observed on the WC linearized PB limit \cite{Kornyshev, Harreis}. At the relative orientation $\varphi_0=\pi$ the helices on different cylinders approach locally to smallest possible separation causing high electrostatic fields that attract counterions more than in other configurations. The interaction in the orientation $\varphi_0=\pi$ is therefore an order of magnitude larger than for $\varphi_0=0$ or $\varphi_0=\pi/2$ at small separations. At large separations the interaction is independent of orientation and specific charge distribution. There it expectantly approaches the uniformly-charged-cylinders results.

At well defined conditions the osmotic counterion contribution can be reduced to such an extent that the electrostatic correlation attraction prevails and pulls the cylinders into a bound state at their closest possible separation, $R=2a$.
Particular for ds-DNA parameters this happens for multivalent counterions, $q\ge 2$, and in defined orientation range $\varphi_0\sim 0.3\pi - 0.6\pi$. The counterions at these conditions are localized away from the interaxial plane and the osmotic contribution is therefore sufficiently reduced for this to happen.

If two helices can freely rotate around their long axes they will settle in the orientation that minimizes their free energy.
At large distances the optimal angle for SC is $\varphi_0=\pi$, which is the most unfavorable configuration in the WC case. Counterions that mediate electrostatic  attraction between both helices tend to orient helices so that the local helical segments approach as much as possible. When surface-to-surface separation is smaller than the counterion diameter, the depletion effect localizes counterions out-of-center and the optimal angle changes.

Our approach in many respects complements the analysis performed by Kornyshev and Leikin \cite{Kornyshev}, which is based on the assumption of complete and ion-specific adsorption of polyvalent counterions onto the surfaces of the helical molecules. Rather, in the case investigated here, the only interaction between the counterions and the macromolecular surface is electrostatic in nature. The fact that even in this case there exists a strong angular dependent attraction between two cylindrical molecules bearing a helical charge motif is quite remarkable in itself, and results of our analysis will certainly be relevant in the case of helical charge distributions where there is no reason to expect any specific non-electrostatic interactions between the counterions and the macromolecular surface.  Further work is in progress with the aim of understanding the effects of finite concentration of monovalent salt, apart from the presence of polyvalent counterions,  on the behavior of this system.

\section{Acknowledgement}
RP and JD would like to acknowledge the financial support by the Slovenian Research Agency under contract
Nr. P1-0055 (Biophysics of Polymers, Membranes, Gels, Colloids and Cells). MK would like to acknowledge the financial support by the Slovenian Research Agency under the young researcher grant.

\end{document}